\documentclass[11pt,a4paper]{article}
\pdfoutput=1

\usepackage{jheppub}
\usepackage{physics}
\usepackage{tensor}
\usepackage{amsmath,amsfonts,amssymb,amsthm}
\usepackage{url}
\usepackage{bm}

\numberwithin{equation}{section}

\newcommand{\fft}[2]{\frac{#1}{#2}}

\newcommand{\nn}{\nonumber}
\newcommand{\abrak}[1]{\left\langle #1 \right\rangle}
\newcommand{\inv}{^{-1}}
\newcommand{\Z}{\mathbb{Z}}
\newcommand{\N}{\mathbb{N}}

\preprint{LCTP-22-16}

\title{Finite $N$ indices and the giant graviton expansion}

\author[a]{James T. Liu}

\emailAdd{jimliu@umich.edu}

\author[a]{and Neville Joshua Rajappa}

\emailAdd{nevjoraj@umich.edu}

\affiliation[a]{Leinweber Center for Theoretical Physics, University of Michigan, Ann Arbor, MI 48109, U.S.A.}

\abstract{The superconformal index of $\mathcal N=4$ super-Yang Mills theory with $U(N)$ gauge group can be written as a matrix integral over the gauge group.  Recently, Murthy demonstrated that this integral can be reexpressed as a sum of terms corresponding to a giant graviton expansion of the index, and provided an explicit formula for the case of a single giant graviton.  Here we give similar explicit formulae for an arbitrary number, $m\ge1$, of giant gravitons.  We provide 1/2 and 1/16 BPS index examples up to the order where three giant gravitons contribute and demonstrate that the expansion of the matrix integral differs from the giant graviton expansion computed in the supergravity dual.  This shows that the giant graviton expansion is not necessarily unique once two or more giant gravitons start appearing.}


\arxivnumber{}

\begin{document}

\maketitle


\section{Introduction}

Motivated by recent developments in supersymmetric gauge theories and AdS/CFT, a lot of attention has been focused on computing the superconformal index in various settings.  One particularly important application has been to the microscopic understanding of AdS black hole entropy as a counting problem via the index in the dual superconformal field theory.  In this setting, one is mainly interested in the large-$N$ limit of the index, which can be explored by various methods.  At the same time, however, the index is well-defined for any finite $N$, and it can be interesting to explore its properties away from the large-$N$ limit.

From a Hamiltonian point of view, the index can be viewed as a refined Witten index
\begin{equation}
Z(\mu_i)=\Tr[(-1)^fe^{-\beta\{\mathcal Q,\mathcal Q^\dagger\}}e^{-\mu_iQ_i}].
\label{eq:index}
\end{equation}
Here $\mathcal Q$ is a supercharge, and the index is refined by the chemical potentials $\mu_i$ corresponding to the charges $Q_i$ that commute with $\mathcal Q$.  Since states with non-zero $\delta\equiv\{\mathcal Q,\mathcal Q^\dagger\}$ cancel in pairs, the index is independent of $\beta$ and essentially counts BPS states preserving some fraction of the supersymmetry.  An important feature of the index is that it is topological and, in many circumstances, can be computed in a weak coupling limit where the theory is more manageable.

Here we focus on the superconformal index of $\mathcal N=4$ super-Yang-Mills with gauge group $U(N)$ \cite{Romelsberger:2005eg,Kinney:2005ej}.  Based on a weak coupling expansion, the index can be obtained by first writing down the single-letter index, $i(\mu_i)$, corresponding to that of a single $\mathcal N=4$ vector multiplet and then performing the Plethystic exponentiation while projecting onto gauge invariant states.  The result takes the schematic form \cite{Kinney:2005ej}
\begin{equation}
    Z_N(q_i)=\int_{U(N)} dU\exp\left[\sum_{k=1}^\infty\fft{i(q_i^k)}k\Tr(U^k)\Tr( U^{\dagger\,k})\right],
\label{eq:ZNq}
\end{equation}
which can be interpreted as a matrix model partition function for $U(N)$ matrices.  Here we have introduced the fugacities $q_i=e^{-\mu_i}$.  The beauty of this expression is that it is valid for any rank $N$.  The single letter index takes different explicit forms depending on what BPS states are being counted and the nature of the refinement by chemical potentials.  Nevertheless, this compact expression is convenient for a systematic investigation of the index at finite $N$.

The 1/16-BPS index for $N\sim\mathcal O(1)$ was studied semi-analytically in \cite{Agarwal:2020zwm,Murthy:2020rbd} where it was noted that the index transitions from multi-graviton behavior to Cardy-like behavior as the charge is increased.  More specifically, the multi-graviton index can be identified with $N=\infty$, where the integral over the gauge group in (\ref{eq:ZNq}), which enforces the gauge singlet condition, is straightforward, with the result
\begin{equation}
    Z_\infty(\mu_i)=\prod_{k=1}^\infty\fft1{1-i(q_i^k)}.
\end{equation}
Equivalently, this result follows because at infinite $N$ there are no trace relations that reduce the number of independent multi-graviton states from the spectrum.

Viewed as a counting problem, the index can be expanded in a power series of the form
\begin{equation}
    Z_N(q_i)=\sum_{\{\ell_i\}} d_N(\ell_i){\textstyle\prod_i} q_i^{\ell_i},
\end{equation}
where the sum is over all vectors of charges corresponding to the $Q_i$ in (\ref{eq:index}).  For small values of charge, $\sum_i\ell_i\lesssim\mathcal O(N)$, the degeneracies $d_N(\ell_i)$ take the universal multi-graviton values $d_\infty(\ell_i)$, while at larger values of charge the degeneracies start to differ \cite{Agarwal:2020zwm,Murthy:2020rbd}.  Technically, this occurs because non-trivial $U(N)$ trace relations show up only at sufficiently high charge of $\mathcal O(N)$.  From an AdS/CFT point of view, this behavior of the finite-$N$ index corresponds to a giant graviton expansion \cite{Arai:2019xmp,Arai:2019wgv,Arai:2019aou,Imamura:2021ytr,Gaiotto:2021xce,Murthy:2022ien,Lee:2022vig}
\begin{equation}
    \fft{Z_N(q_i)}{Z_\infty(q_i)}=1+\sum_{m=1}^\infty G_N^{(m)}(q_i),
\label{eq:gge}
\end{equation}
where $G_N^{(m)}(q_i)$ is the contribution of $m$ giant gravitons to the index.  Since the contribution of a single giant graviton operator scales as $\mathcal O(N)$, the contribution $G_N^{(m)}(q_i)$ scales as $q^{\mathcal O(mN)}$, with the precise scaling governed by the behavior of the single letter index $i(q_i)$.

The giant graviton expansion, (\ref{eq:gge}), is quite remarkable, as it relates a purely field theoretic calculation of the $\mathcal N=4$ SYM index to its AdS/CFT dual where the states are apparently fully encoded by combining Kaluza-Klein gravitons on AdS$_5\times S^5$ that generate the multi-graviton index $Z_\infty(q_i)$ with wrapped D3-branes, namely the giant gravitons, that account for the $G_N^{(m)}(q_i)$ contributions.  Moreover, as conjectured in \cite{Gaiotto:2021xce}, in many (but not all) cases, if we focus on the scaling behavior of a single fugacity, the giant graviton contribution takes on a universal form
\begin{equation}
    G_N^{(m)}(q;y)=q^{mN}\hat G^{(m)}(q;y),
\end{equation}
where $y$ denotes the remaining fugacities that are being held fixed.  This expansion is universal in the sense that $\hat G^{(m)}(q;y)$ is independent of the rank, $N$, of the gauge group.

The expansion, (\ref{eq:gge}), and in particular the contributions $G_N^{(m)}(q_i)$, was derived directly in \cite{Arai:2019xmp,Imamura:2021ytr} from the gauge theory on the wrapped D3-branes, however with some assumptions on handling multiply wrapped D3-branes.  Subsequently, a separate treatment of this expansion was given by Murthy in \cite{Murthy:2022ien} by working directly with the matrix integral form of the index, (\ref{eq:ZNq}), and re-expressing it in terms of a system of free fermions.  In the matrix model, the giant graviton contribution, $G_N^{(m)}(q_i)$, then takes the form of a fermionic determinant
\begin{equation}
    G_N^{(m)}(q_i)\sim\det(\tilde K(r_i,r_j)),
\end{equation}
where $\tilde K(r_i,r_j)$ is a fermion two-point function.  This expression is written rather schematically, but will be made more precise in the following section.

For the contribution of a single giant graviton, the determinant is trivial, and Murthy provides an explicit formula for $G_N^{(1)}(q_i)$, as well as examples for the $1/2$-BPS, $1/8$-BPS and $1/16$-BPS indices \cite{Murthy:2022ien}.  In this paper, we extend the results of \cite{Murthy:2022ien} by providing explicit formulas for any number, $m$, of giant gravitons.  These resulting expressions for $G_N^{(m)}(q_i)$ can be evaluated by computer algebra, and we provide several examples for the $1/2$ and $1/16$-BPS index.  By explicit computation, we demonstrate that the matrix integral expansion of \cite{Murthy:2022ien} differs from the wrapped D3-brane expansion of \cite{Imamura:2021ytr} at the order when two giant gravitons, namely $G_N^{(2)}$, start contributing.  While both expansions sum to the index, $Z_N(q_i)$, the individual contributions differ, even for the single giant graviton term, $G_N^{(1)}$.  This raises some questions about the physical significance of the giant graviton expansion and whether it is possible to map from one expansion to the other.

While the expansions of \cite{Murthy:2022ien} and \cite{Imamura:2021ytr} differ in their explicit form, we can see that both agree on the order of the giant graviton expansions.  In particular, in \cite{Murthy:2022ien}, it was shown that if we focus on one overall fugacity with single letter index $i(q)$ of order $\alpha$, then the order of $G_N^{(m)}(q)$ is at least $\alpha(N+1)$ for $m=1$ or $\alpha(mN+m(m+1)/2+1)$ for $m\ge2$.  Here the order of a function of $q$ denotes the lowest power of $q$ that shows up in its Taylor expansion.  While this matches the physical notion of a giant graviton expansion, it was noted that this is not necessarily an optimal bound.  Given the explicit form of the multi-giant graviton index, we improve this bound by demonstrating that $G_N^{(m)}(q)$ has order at least $\alpha m(N+m)$, and that this bound is saturated for generic single letter indices $i(q)$ of order $\alpha$, which is in agreement with the wrapped D3-brane calculation of \cite{Imamura:2021ytr}.

The outline of this paper is as follows.  In the next section, we first summarize the giant graviton expansion of \cite{Murthy:2022ien}, and then provide an explicit formula for the $m$ giant graviton contribution $G_N^{(m)}$.  We also show that this expansion starts at order $\alpha m(N+m)$, as indicated above.  We then present examples for the 1/2 BPS and 1/16 BPS indices of $\mathcal N=4$ SYM in section~\ref{sec:examples} and highlight that the expansion is not unique.  Finally, we make a few concluding remarks in section~\ref{sec:discussion}.  Some of the technical manipulations are relegated to two appendices.

\section{The giant graviton expansion}
\label{sec:gge}

In \cite{Murthy:2022ien}, Murthy provides an elegant determinant expression for the contribution $G_N^{(m)}(q_i)$ of $m$ giant gravitons.  Here we summarize the result of \cite{Murthy:2022ien}, following the same notation with only minor changes.  While we are mainly interested in the superconformal index, (\ref{eq:ZNq}), the starting point of \cite{Murthy:2022ien} is the more general matrix integral
\begin{equation}
    Z_N(\mathbf g)=\int_{U(N)} dU\exp\left[\sum_{k=1}^\infty\fft{g_k}k\Tr(U^k)\Tr( U^{\dagger\,k})\right],
\label{eq:matint}
\end{equation}
where the $g_k$ are viewed as a set of couplings.  The index is then obtained by taking $g_k=i(q_i^k)$, corresponding to Plethystic exponentiation of the single letter index.

The matrix integral, (\ref{eq:matint}), then admits the expansion \cite{Murthy:2022ien}
\begin{equation} \label{eq:ggexp}
    \fft{Z_N(\mathbf g)}{Z_\infty(\mathbf g)}=1+\sum_{m=1}^\infty G_N^{(m)}(\mathbf g),
\end{equation}
where
\begin{equation}
    Z_\infty(\mathbf g)=\prod_{k=1}^\infty\fft1{1-g_k},
\end{equation}
and
\begin{equation}
\label{eq:gm}
    G_N^{(m)}(\mathbf{g}) = (-1)^m \sum_{ \substack{N < r_1 < \dots < r_m \\ r_i \in \mathbb{Z} + \frac{1}{2}} } K_g^{(m)}(r_1, \dots, r_m).
\end{equation}
Here $K_g^{(m)}(r_1, \dots, r_m)$ is related to the $\langle\cdots\rangle_\mathbf g$ averaging of the determinant of a fermion two-point function $\tilde K(r_i, r_j)$, namely
\begin{equation} \label{eq:Kgmdet}
    K_g^{(m)}(r_1, \dots, r_m) = \frac{1}{Z_\infty(\textbf{g})} \abrak{\tilde Z_\infty \det(\tilde K(r_i, r_j))_{i, j = 1}^m}_\mathbf{g}.
\end{equation}
The averaging acts on a power series $f(\mathbf t^+,\mathbf t^-)$ according to
\begin{equation} \label{eq:averaging}
    \langle f\rangle_{\mathbf g}=\prod_{k=1}^\infty\int\fft{dt_k^+dt_k^-}{2\pi kg_k}e^{-t_k^+t_k^-/kg_k}f(\mathbf t^+,\mathbf t^-).
\end{equation}
The expressions inside the averaging in (\ref{eq:Kgmdet}) are implicitly functions of $\mathbf t^+$ and $\mathbf t^-$.  In particular,
\begin{equation}
    \tilde Z_\infty=\exp(\sum_{k=1}^\infty\fft{t_k^+t_k^-}k),
\end{equation}
and the two-point function $\tilde K(r,s)$ is defined via its generating function
\begin{equation} \label{eq:2ptgen}
    \tilde\kappa(z,w)=\sum_{r,s\in\mathbb Z+\fft12}\tilde K(r,s)z^rw^{-s}=\fft{\sqrt{zw}}{z-w}\exp(\sum_{k=1}^\infty\left( t_k^+(z^k-w^k)-t_k^-(z^{-k}-w^{-k})\right)).
\end{equation}
The above expressions are all that are necessary to evaluate the giant graviton index.  However, their derivation is quite elegant and well explained in \cite{Murthy:2022ien}, to which we refer the reader for a more thorough discussion.

\subsection{The contribution of one giant graviton}
\label{sec:gge1}

For the single giant graviton contribution, (\ref{eq:gm}) and (\ref{eq:Kgmdet}) reduces to
\begin{equation}
    G_N^{(1)}(\mathbf g)=-\sum_{ N < r,~r \in \mathbb{Z} + \fft12}\fft1{Z_\infty(\mathbf g)}\langle\tilde Z_\infty\tilde K(r,r)\rangle_{\mathbf g}.
\label{eq:GN1giant}
\end{equation}
To evaluate
this expression, we start with the $\langle\cdots\rangle_\mathbf g$ average of the generating function
\begin{equation}
    \langle\tilde Z_\infty\tilde\kappa(z,w)\rangle_{\mathbf g}=Z_\infty(\mathbf g)\fft{\sqrt{w/z}}{1-w/z}\exp(-\sum_{k=1}^\infty\fft{\gamma_k}k(z^k-w^k)(z^{-k}-w^{-k})),
\label{eq:gavegf}
\end{equation}
where
\begin{equation}
    \gamma_k=\fft{g_k}{1-g_k}.
\label{eq:gammak}
\end{equation}
We then transform back to $\langle\tilde Z_\infty\tilde K(r,s)\rangle_{\mathbf g}$ using the formal inversion formula
\begin{equation}
    \tilde K(r,s)=\oint\fft{dz}{2\pi iz}\oint\fft{dw}{2\pi iw}\tilde\kappa(z,w)z^{-r}w^s.
\end{equation}
Setting $s=r$ and performing the sum over $r$ in (\ref{eq:GN1giant}) then gives
\begin{align}
    G_N^{(1)}(\mathbf g)&=-\oint\fft{dz}{2\pi iz}\oint\fft{dw}{2\pi iw}\fft{(w/z)^{N+1/2}}{1-w/z}\fft1{Z_\infty(\mathbf g)}\langle\tilde Z_\infty\tilde\kappa(z,w)\rangle_{\mathbf g}\nn\\
    &=-\oint\fft{dz}{2\pi iz}\oint\fft{dw}{2\pi iw}\fft{(w/z)^{N+1}}{(1-w/z)^2}\exp(-\sum_{k=1}^\infty\fft{\gamma_k}k(z^k-w^k)(z^{-k}-w^{-k})),
\end{align}
where the second line is obtained by substituting in (\ref{eq:gavegf}).  Since the integrand is a function of $w/z$, we can make the substitution $\zeta=w/z$ and trivially perform one of the contour integrals to obtain
\begin{equation}
    G_N^{(1)}(\mathbf g)=-\oint\fft{d\zeta}{2\pi i\zeta}\fft{\zeta^{N+1}}{(1-\zeta)^2}\exp(-\sum_{k=1}^\infty\fft{\gamma_k}k(1-\zeta^k)(1-\zeta^{-k})),
\end{equation}
or equivalently \cite{Murthy:2022ien}
\begin{equation}
    G_N^{(1)}(\mathbf g)= \left.\frac{-\zeta}{(1 - \zeta)^2} \exp \left( -\sum_{k = 1}^\infty \frac{\gamma_k}{k} (1-\zeta^k)(1- \zeta^{-k}) \right)\right|_{\zeta^{-N}},
\end{equation}
where $f(\zeta)|_{\zeta^{-N}}$ indicates we are to take the coefficient of $\zeta^{-N}$ in the series expansion of $f(\zeta)$ for $\zeta$ around zero.  As noted in \cite{Murthy:2022ien}, this expression is still somewhat formal; since the series includes all negative and positive powers of $\zeta$, it is rather ill behaved.  To make sense of the expression, we must first expand in $\gamma_k$ and then, at fixed order in $\gamma_k$, expand as a Laurent series in $\zeta$.

\subsection{The contribution of multiple giant gravitons}
\label{sec:ggem}

Having reviewed the process given in \cite{Murthy:2022ien} to obtain the contribution of a single giant graviton, we now turn to the contribution of $m$ giant gravitons with $m\ge1$.  Recalling that this contribution is given by (\ref{eq:gm}), the general procedure is to evaluate the determinant in (\ref{eq:Kgmdet}) using the generating function, (\ref{eq:2ptgen}).  However, before we do this, we first make a technical simplification by changing the ordered sum in (\ref{eq:gm}) to an unrestricted sum by making use of the properties of the determinant.

The key observation is that, since $K_g^{(m)}(r_1,\ldots,r_m)$ is a determinant, it satisfies several basic properties related to the permutations%
\footnote{For permutations, we use the following notation.  Let $\sigma \in S_m$ be a permutation. We then denote the number $\sigma$ sends $i$ to by $\sigma_i$. If $\sigma$ is thought of as a bijection, then this notation is just $\sigma_i \equiv \sigma(i)$.}
of the $r_i$.  We first note that in \eqref{eq:Kgmdet} the $i,j$ indices are the indices of the matrix of which we take the determinant. Thus, applying any permutation $\tau \in S_m$ on the variables $r_1, \dots, r_m$ effectively corresponds to permuting both the rows and the columns of the matrix by $\tau$. Since the determinant is invariant under the simultaneous permutation of the rows and columns, we conclude that
\begin{equation}
    K_g^{(m)}(r_{\tau_1}, \dots, r_{\tau_m}) = K_g^{(m)}(r_1, \dots, r_m).
\end{equation}
By the same argument, setting any two $r_i, r_j$ to be equal results in the determinant being zero, since the matrix then has two columns (and two rows) equal to each other.

Assuming $r_i \in \mathbb{Z} + 1/2$ for brevity, and using these observations, we see that 
\begin{align}
    \sum_{N < r_i} K_g^{(m)}(r_1, \dots, r_m) &= \sum_{\sigma \in S_m}  \sum_{N < r_{\sigma_1} < \dots < r_{\sigma_m}} K_g^{(m)}(r_1, \dots, r_m) + \sum_{i \neq j} \sum_{r_i = r_j} K_g^{(m)}(r_1, \dots, r_m)\nn \\
    &= \sum_{\sigma \in S_m}  \sum_{N < r_1 < \dots < r_m} K_g^{(m)}(r_{{\sigma\inv}_1}, \dots, r_{{\sigma\inv}_m}) +0\nn\\
    &= \sum_{\sigma \in S_m}  \sum_{N < r_1 < \dots < r_m} K_g^{(m)}(r_1, \dots, r_m) \nn\\
    &= m!  \sum_{N < r_1 < \dots < r_m} K_g^{(m)}(r_1, \dots, r_m).
\end{align}
Substituting this into (\ref{eq:gm}), we thus arrive at
\begin{equation} \label{eq:newGm}
     G_N^{(m)}(\textbf{g}) = \frac{(-1)^m}{m!} \sum_{N < r_i,~r_i \in \Z + \frac{1}{2}} K_g^{(m)}(r_1, \dots, r_m).
\end{equation}
As we shall see below, this is useful because it is much easier to deal with the unrestricted sum than the restricted sum.

We now proceed to evaluate $K_g^{(m)}(r_1,\ldots,r_m)$ defined in (\ref{eq:Kgmdet}) by making use of the generating function for $\tilde K(r_i,r_j)$ given in (\ref{eq:2ptgen}).  Using the permutation expression of the determinant, we can rewrite (\ref{eq:Kgmdet}) as
\begin{equation} \label{eq:Kgm}
    \begin{aligned}
        K_g^{(m)}(r_1, \dots, r_m) &= \frac{1}{Z_\infty(\textbf{g})} \sum_{\sigma \in S_m} (-1)^\sigma \abrak{\tilde Z_\infty \prod_{i = 1}^m \tilde K(r_i, r_{\sigma_i}) }_\textbf{g}
    \end{aligned}
\end{equation}
Since $\tilde K(r,s)$ is defined through its generating function $\tilde\kappa(z,w)$ in (\ref{eq:2ptgen}), we can evaluate the more general product
\begin{equation} \label{eq:risi}
    \abrak{\tilde Z_\infty \prod_{i = 1}^m \tilde K(r_i, s_i) }_\textbf{g}
\end{equation}
by use of a multidimensional generating function. To this end, we define the generating function of \eqref{eq:risi} as
\begin{align} \label{eq:genF}
    F(w_i, z_i) &:= \sum_{r_i, s_i \in \Z + \frac{1}{2}} \prod_{i = 1}^m z_i^{r_i} w_i^{-s_i} \abrak{\tilde Z_\infty \prod_{i = 1}^m \tilde K(r_i, s_i)}_\textbf{g} 
    = \abrak{\tilde Z_\infty \prod_{i = 1}^m \tilde \kappa(w_i, z_i)}_\textbf{g}.
\end{align}
Inverting this relation allows us to write
\begin{equation}
    \abrak{\tilde Z_\infty \prod_{i = 1}^m \tilde K(r_i, s_i)}_\textbf{g} = \left[ F(w_i, z_i) \right]_{ w_i^{-s_i}z_i^{r_i}},
\end{equation}
where the notation, as indicated above, is to take the corresponding powers of $w_i$ and $z_i$ in the series expansion about the origin.  We thus have
\begin{equation} \label{eq:newKg}
    K_g^{(m)}(r_1, \dots, r_m) = \frac{1}{Z_\infty(\textbf{g})} \sum_{\sigma \in S_m} (-1)^\sigma \left[ F(w_i, z_i) \right]_{w_i^{-r_{\sigma_i}} z_i^{r_i}}.
\end{equation}

We can compute the generating function $[F(w_i, z_i)]$ by substituting in the definition of $\kappa(w_i, z_i)$ in \eqref{eq:2ptgen} and working out the averaging \eqref{eq:averaging}.  The manipulations are straightforward, and the result is
\begin{equation}
    F(w_i,z_i)=Z_\infty(\textbf{g}) \prod_{i = 1}^m \frac{\sqrt{z_i w_i}}{z_i - w_i} \exp \left( - \sum_{k = 1}^\infty \frac{\gamma_k}{k} \alpha_k \beta_k \right),
\label{eq:Fwizi}
\end{equation}
where
\begin{align} \label{eq:alpbetm}
    \alpha_k &= \sum_{i = 1}^m z_i^k - w_i^k, &
    \beta_k &= \sum_{i = 1}^m z_i^{-k} - w_i^{-k},
\end{align}
and $\gamma_k$ is defined in (\ref{eq:gammak}).  Note that this is a straightforward generalization of (\ref{eq:gavegf}) to the multidimensional case.

In order to obtain $K_g^{(m)}(r_1, \dots, r_m)$, we need to extract the coefficient of $w_i^{-r_{\sigma_i}} z_i^{r_i}$ from $F(w_i,z_i)$.  Since this is simply taking a particular coefficient of the corresponding power series expansion, the variables $w_i$ and $z_i$ are dummy variables.  Thus, we may make the replacement $w_i\to w_{\sigma_i}$ to write
\begin{equation}
    \left[ F(w_i, z_i) \right]_{w_i^{-r_{\sigma_i}} z_i^{r_i}}=\left[ F(w_{\sigma_i}, z_i) \right]_{w_i^{-r_i} z_i^{r_i}}.
\end{equation}
Noting that $\alpha_k$ and $\beta_k$, as well as the product $\prod\sqrt{z_iw_i}$, in (\ref{eq:Fwizi}) are permutation invariant, we then have
\begin{equation} \label{eq:geneqrs}
    [F(w_i, z_i)]_{w_i^{-r_{\sigma_i}}z_i^{r_i}} = \left[ Z_\infty(\textbf{g}) \prod_{i = 1}^m \frac{\sqrt{z_i w_i}}{z_i - w_{\sigma_i}} \exp \left( - \sum_{k = 1}^\infty \frac{\gamma_k}{k} \alpha_k \beta_k \right) \right]_{w_i^{-r_i} z_i^{r_i}}.
\end{equation}
In particular, note that the only term that is sensitive to the permutation $\sigma$ is the denominator term $z_i-w_{\sigma_i}$.  Thus, when we insert this into \eqref{eq:newKg}, we only need to evaluate
\begin{equation}
    \sum_{\sigma \in S_m}(-1)^\sigma\prod_{i=1}^m\fft1{z_i-w_{\sigma_i}}=\prod_{k=1}^m{}\fft1{z_k}\det\left( \frac{1}{1 - w_j/z_i} \right)_{i, j = 1}^m.
\end{equation}
The result is then
\begin{equation}
    K_g^{(m)}(r_1, \dots, r_m) = \left[ \prod_{k = 1}^m \sqrt{\fft{w_k}{z_k}} \det \left( \frac{1}{1 - w_j/z_i} \right)_{i, j = 1}^m \exp \left( - \sum_{k = 1}^\infty \frac{\gamma_k}{k} \alpha_k \beta_k \right) \right]_{w_i^{-r_i} z_i^{r_i}},
\end{equation}
which can be written formally as the inverse transform
\begin{equation}
    K_g^{(m)}(r_1,\ldots,r_m)=\prod_{i=1}^m\oint\fft{dz_i}{2\pi iz_i}\fft{dw_i}{2\pi iw_i}\left(\fft{w_i}{z_i}\right)^{r_i}K_g^{(m)}(w_i,z_i),
\label{eq:cintexp}
\end{equation}
where
\begin{equation}
    K_g^{(m)}(w_i,z_i)=\prod_{k = 1}^m \sqrt{\fft{w_k}{z_k}} \det \left( \frac{1}{1 - w_j/z_i} \right)_{i, j = 1}^m \exp \left( - \sum_{k = 1}^\infty \frac{\gamma_k}{k} \alpha_k \beta_k \right).
\end{equation}
We have thus worked out an explicit form for the object $K_g^{(m)}(r_1, \dots, r_m)$, which is what we set out to do.

Finally, we can sum over the half-integers $r_i$ to obtain the $m$ giant graviton contribution $G_N^{(m)}$ as given in (\ref{eq:newGm}).  Performing the sum over $r_i$ inside the integrals of (\ref{eq:cintexp}) gives
\begin{align}
    G_N^{(m)}(\mathbf g)&=\fft{(-1)^m}{m!}\prod_{i=1}^m\oint\fft{dz_i}{2\pi iz_i}\fft{dw_i}{2\pi iw_i}\fft{(w_i/z_i)^{N+1/2}}{1-w_i/z_i}K_g^{(m)}(w_i,z_i)\nn\\
    &=\fft{(-1)^m}{m!}\prod_{i=1}^m\oint\fft{dz_i}{2\pi iz_i}\fft{dw_i}{2\pi iw_i}\fft{(w_i/z_i)^{N+1}}{1-w_i/z_i}\det \left( \frac{1}{1 - w_j/z_i} \right)_{i, j = 1}^m \exp \left( - \sum_{k = 1}^\infty \frac{\gamma_k}{k} \alpha_k \beta_k \right).
\end{align}
Since the contour integrals pick out the appropriate residues at the origin, we finally arrive at the contribution of $m$ giant gravitons
\begin{equation}
    G_N^{(m)}(\textbf{g)} = \left[ \frac{(-1)^m}{m!} \left( \prod_{i = 1}^m \frac{w_i/z_i}{1 - w_i/z_i} \right) \det \left( \frac{1}{1 - w_j/z_i} \right)_{i, j = 1}^m \exp \left( - \sum_{k = 1}^\infty \frac{\gamma_k}{k} \alpha_k \beta_k \right) \right]_{w_i^{-N} z_i^{N}},
\label{eq:GmNfinal}
\end{equation}
where $\{\alpha_k, \beta_k\}$ and $\gamma_k$ are defined in (\ref{eq:alpbetm}) and (\ref{eq:gammak}), respectively, which we repeat here for convenience
\begin{align}
    \alpha_k &= \sum_{i = 1}^m z_i^k - w_i^k, &
    \beta_k &= \sum_{i = 1}^m z_i^{-k} - w_i^{-k},
    &
    \gamma_k &= \frac{g_k}{1 - g_k}.
\end{align}
Recall that $g_k$ are the general couplings in the matrix integral (\ref{eq:matint}), and can be replaced by the single letter index $i(q_i^k)$ when computing the superconformal index.  This generalizes the explicit formula for $G_N^{(1)}(\mathbf g)$ presented in \cite{Murthy:2022ien}.

\subsection{The order of the $m$ giant graviton contribution}
\label{sec:ggeorder}

When applied to the superconformal index, the couplings $g_k$ are related to the single letter index $i(q_i)$ according to $g_k=i(q_i^k)$.  Taking only a single fugacity $q$ for simplicity, $i(q)$ can be expanded as a series in $q$.  Assuming the leading term in the expansion is $i(q)=q^\alpha+\cdots$, it is easy to see that $\gamma_k=q^{\alpha k}+\cdots$.  Here $\alpha$ denotes the order of the single letter index $i(q)$.  So long as $\gamma_k$ is of minimum order $\alpha k$, it was shown in \cite{Murthy:2022ien} that the contribution $G_N^{(m)}(\mathbf g)$ of $m$ giant gravitons is of order at least $\alpha(mN+m(m+1)/2)$.  The way this comes about is that the determinant term in \eqref{eq:GmNfinal} gives a sort of ``determinant structure" on the series terms, and this structure results in the cancellation of terms of low order, leading to this lower bound.

Empirically, we find that while the contribution of one giant graviton saturates the bound $\alpha(mN+m(m+1)/2)$, the contribution of two or more giant gravitons actually starts at the higher order of $\alpha m(N+m)$.  This suggests that there are additional cancellations present in (\ref{eq:GmNfinal}) as a result of the determinant, which we now investigate.  In the next section, we show by examples that this bound is saturated, implying that the order of $G^{(m)}_N(\mathbf{g})$ is in fact exactly $\alpha m(N + m)$, at least for the generic case.

Our proof goes roughly as follows: we first expand the determinant term in \eqref{eq:GmNfinal} and confirm that the obvious cancellations leads to the original lower bound of $\alpha(mN+m(m+1)/2)$ given in \cite{Murthy:2022ien}.  However, we show that when we actually select the $w_i^{-N} z_i^N$ terms to compute the contribution of $m$ giant gravitons in \eqref{eq:GmNfinal}, a second set of cancellations arises, in a very similar flavour to the ones we just described. These effectively result from the determinant structure coupling with the rest of the equation. In particular, it uses the terms that come from the prefactor to the determinant in \eqref{eq:GmNfinal}, the invariance of the exponential term with respect to independent permutations of $w_i$ and $z_i$, and the invariance of the order in which we select the $w_i^{-N} z_i^N$ terms, to create a similar setup as in the previous set of cancellations.

As a side note, these cancellations also have the effect of introducing \textit{new} permutation structures on the remaining terms. They are not essential to this proof, since the same result can be achieved without the use of these structures. However, we do think that these structures are noteworthy and could have further implications for the giant graviton expansion.

\subsubsection{Proof of determinant cancellations}

We start with the determinant part of $G_N^{(m)}(\mathbf g)$ given in \eqref{eq:GmNfinal}. Using the permutation form of the determinant and then expanding it into a power series (which is valid since we assume $|w_j/z_i| < 1$) gives us
\begin{equation} \label{eq:detterm}
    \det \left( \frac{1}{1 - w_j/z_i} \right)_{i, j = 1}^m = \sum_{k_i = 0}^\infty \left( \sum_{\sigma \in S_m} (-1)^\sigma \prod_{i = 1}^m \left( \frac{w_{\sigma_i}}{z_i} \right)^{k_i} \right).
\end{equation}
We see that this is a sum of an infinite number of terms of the form
\begin{equation} \label{eq:detterms}
    P(k_1, \dots, k_m) := \sum_{\sigma \in S_m} (-1)^\sigma \prod_{i = 1}^m \left( \frac{w_{\sigma_i}}{z_i} \right)^{k_i}
    = \det \left( \left( \frac{w_j}{z_i} \right)^{k_i} \right)_{i, j = 1}^m,
\end{equation}
where the $k_i$ are nonnegative integers. However, not all of these terms survive. To see this, we rewrite \eqref{eq:detterms} by pulling the $z_i$'s out of the determinant:
\begin{equation} \label{eq:pterms}
    P(k_1, \dots, k_m) = \prod_{i = 1}^m z_i^{-k_i} \cdot \det(w_j^{k_i})_{i, j = 1}^m.
\end{equation}
Now, if the $k_i$ are not distinct, then two of the rows of the matrix would be equal to each other, and thus its determinant would vanish.

Hence, for \eqref{eq:detterms} to be nonvanishing, the $k_i$'s must be distinct. This introduces a sort of \textit{permutation} structure onto the $P(k_1, \dots, k_m)$ in the following way: if we have a fixed set of $m$ distinct non-negative integers to pick the $k_i$'s from, the possible values of $(k_1, \dots, k_m)$ correspond exactly to the permutations of this set of integers. We can use this to define
\begin{equation}
    P(A, \tau) := P(A_{\tau_1}, \dots, A_{\tau_m})
    = P(\tau^A_i, \dots, \tau^A_m),
\end{equation}
where we define%
\footnote{This $\tau_i^A$ notation may seem a bit unusual, but its goal is to emphasize the permutation aspect of the term. However, note that we will use both notations interchangeably to emphasize the aspect that is being worked with.}
$\tau^A_i := A_{\tau_i}$. We now rewrite \eqref{eq:detterm} as
\begin{equation} \label{eq:detexpr}
    \det \left( \frac{1}{1 - w_j/z_i} \right)_{i, j = 1}^m = \sum_{\substack{A \in 2^{\N \cup \{0\}} \\ |A| = m}} \sum_{\tau \in S_m} P(A, \tau),
\end{equation}
where $|A| = m$ means that there are $m$ elements in the set $A$.

We now incorporate our result into the giant graviton expression \eqref{eq:GmNfinal} as a whole. We expand the $(w_i/z_i)/(1 - w_i/z_i)$ prefactor as a power series in $w_i$ and $z_i$ and make some convenient rearrangement of the terms; this computation is a bit tedious, and the details are relegated to Appendix \ref{app:gmnrew}. The end result is that
\begin{equation} \label{eq:Kmint1}
    \hat K_m(\textbf{g}) = \frac{(-1)^m}{m!} \sum_{\substack{A \in 2^\N \\ |A| = m}} \sum_{\ell_j = 0}^\infty \left( \prod_{j = 1}^m \left( \frac{w_j}{z_j} \right)^{\ell_j} \sum_{\tau \in S_m} P(A, \tau) \right) \cdot E(w_k, z_k),
\end{equation}
where
\begin{equation}
    E(w_k, z_k)=\exp \left( - \sum_{k = 1}^\infty \frac{\gamma_k}{k} \alpha_k \beta_k \right),
\label{eq:Edef}
\end{equation}
Here $\hat K_m(\textbf{g})$ is the generating function of $G_N^{(m)}(\mathbf g)$ in the sense that
\begin{equation}
    G_N^{(m)}(\mathbf g)=\left[\hat K_m(\textbf{g})\right]_{{w_i^{-N}z_i^{N}}}.
\label{eq:GfromK}
\end{equation}
We now work with the terms of the form
\begin{equation} \label{eq:qterms}
    Q(A, \ell_1, \dots, \ell_m) := \prod_{j = 1}^m \left( \frac{w_j}{z_j} \right)^{\ell_j} \sum_{\tau \in S_m} P(A, \tau),
\end{equation}
in the expression for $\hat K_m(\textbf{g})$.  Note that here the $\ell_j$ are nonnegative integers.  Expanding $P(A, \tau)$ via its definition in \eqref{eq:pterms} and rearranging terms, we see that
\begin{align}
    Q(A, \ell_1, \dots, \ell_m) &= \prod_{j = 1}^m \left( \frac{w_j}{z_j} \right)^{\ell_j} \left( \sum_{\tau^A, \sigma \in S_m} (-1)^\sigma \prod_{i = 1}^m \left( \frac{w_{\sigma_i}}{z_i} \right)^{\tau^{A}_i} \right) \nn\\
    &= \sum_{\tau^A, \sigma \in S_m} (-1)^\sigma \prod_{i = 1}^m w_i^{(\tau \circ \sigma\inv)^A_i + \ell_i} \cdot \prod_{i = 1}^m z_i^{- \tau^{A}_i - \ell_i}
\end{align}
Note that in the result above, the relevant quantity associated to the permutation $\sigma$ is $\alpha := \tau \circ \sigma\inv$. This freely varies over $S_m$ since $\sigma$ and $\tau$ freely vary over $S_m$, and thus, we may replace the sum over $\sigma$ with a sum over $\alpha$ to get that
\begin{align} \label{eq:qexplicit}
    Q(A, \ell_1, \dots, \ell_m) &= \sum_{\alpha^A, \tau^A \in S_m} (-1)^{(\alpha\inv \circ \tau)^A} \prod_{i = 1}^m w_i^{\alpha^A_i + \ell_i} \cdot \prod_{i = 1}^m z_i^{- \tau^{A}_i - \ell_i} \nn \\
    &= \det \left( z_i^{-(A_j + \ell_i)} \right)_{i, j = 1}^m \det \left( w_i^{A_j + \ell_i} \right)_{i, j = 1}^m.
\end{align}
We thus see the emergence of a second determinant, and just like before this will lead to a similar set of cancellations.

Recalling that $\hat K_m(\textbf{g})$ is the generating function for $G^{(m)}_N(\textbf{g})$, we can now write down a compact expression for the giant graviton contribution in terms of $Q(A,\ell_1,\ldots,\ell_m)$.  In particular, we combine the expression for $\hat K_m(\mathbf g)$ in \eqref{eq:Kmint1} with the notation introduced in \eqref{eq:qterms} to arrive at
\begin{equation} \label{eq:Gmnint1}
    G^{(m)}_N(\textbf{g}) = \frac{(-1)^m}{m!} \sum_{\substack{A \in 2^\N \\ |A| = m}} \sum_{\ell_j = 0}^\infty \Bigl[ Q(A, \ell_1, \dots, \ell_m) E(w_k, z_k) \Bigr]_{w_i^{-N} z_i^N}.
\end{equation}
We now show that the term $\left[ Q(A, \ell_1, \dots, \ell_m) E(w_k, z_k) \right]_{w_i^{-N} z_i^N}$ vanishes if the $\ell_i$ are not distinct.

Suppose the $\ell_i$ were not distinct, i.e., there exists $\ell_a = \ell_b$ with $a \neq b$. Then, let $\rho \in S_m$ be the permutation that only exchanges $a$ and $b$ and leaves everything else intact.  We note that $E(w_k, z_k)$ is permutation invariant under independent permutations of $w_i$ and $z_i$ so that, in particular, $E(w_k, z_k) = E(w_{\rho_k}, z_k)$. Also by virtue of the definition of $\rho$, we have that $(\ell_1, \dots, \ell_m) = (\ell_{\rho_1}, \dots, \ell_{\rho_m})$. From these observations, we have that
\begin{equation} \label{eq:cancel2p2}
    \Bigl[ Q(A, \ell_1, \dots, \ell_m) E(w_k, z_k) \Bigr]_{w_i^{-N} z_i^N} 
    = \Bigl[ Q(A, \ell_{\rho_1}, \dots, \ell_{\rho_m}) E(w_{\rho_k}, z_k) \Bigr]_{w_i^{-N} z_i^N}.
\end{equation}
Let us investigate the right-hand side of this expression. Expanding the $Q$ term via the result of \eqref{eq:qexplicit}, it becomes
\begin{equation} \label{eq:cancel2p3}
    \left[ \det \left( z_i^{-(A_j + \ell_{\rho_i})} \right)_{i, j = 1}^m \det \left( w_i^{A_j + \ell_{\rho_i}} \right)_{i, j = 1}^m E(w_{\rho_k}, z_k) \right]_{w_i^{-N} z_i^N}.
\end{equation}
We now make three subtle but important changes to the above term:
\begin{enumerate}
    \item We remove the permutation $\rho$ from the $z_i$ determinant, since $\ell_{\rho_i} = \ell_i$ by the definition of $\rho$. We do this because we want the permutation to act only on the $w_i$.
    
    \item In the $w_i$ determinant, we permute the rows around by $\rho$, which introduces a factor of $(-1)^\rho$.
    
    \item We also change the order in which we take the coefficients from $w_i^{-N}$ to $w_{\rho_i}^{-N}$.
\end{enumerate}
As a result of these, \eqref{eq:cancel2p3} becomes
\begin{equation}
    \left[ \det \left( z_i^{-(A_j + \ell_i)} \right)_{i, j = 1}^m \cdot (-1)^\rho \det \left( w_{\rho_i}^{A_j + \ell_i} \right)_{i, j = 1}^m E(w_{\rho_k}, z_k) \right]_{w_{\rho_i}^{-N} z_i^N}
\end{equation}
and making the replacement $w_i \to w_{\rho\inv_i}$ (which is fine since they are dummy variables), it turns into
\begin{equation}
    - \Bigl[Q(A, \ell_1, \dots, \ell_m) E(w_k, z_k)\Bigr]_{w_i^{-N} z_i^N},
\end{equation}
where we used the fact that $\rho$, which performs a single exchange, is an odd permutation.  Thus, we have that $[Q(A, \ell_1, \dots, \ell_m) E(w_k, z_k)]_{w_i^{-N} z_i^N}$ is negative of itself, and must hence be zero.

We have thus proven that only the terms that have distinct $\ell_i$ contribute to $G_N^{(m)}(\mathbf g)$. Since we are dealing with distinct sets of non-negative integers once again, we may introduce another permutation structure
\begin{equation}
    Q(B, A, \alpha) := Q(A, \alpha_1, \dots, \alpha_m)
\end{equation}
and finally express \eqref{eq:GmNfinal} as
\begin{equation} \label{eq:GmnQ}
    G^{(m)}_N(\textbf{g}) = \frac{(-1)^m}{m!} \sum_{\substack{A \in 2^\N, B \in 2^{\N \cup \{ 0 \}} \\ |A| = |B| = m}} \sum_{\alpha \in S_m} \Bigl[ Q(B, A, \alpha) E(w_k, z_k) \Bigr]_{w_i^{-N} z_i^N}.
\end{equation}
Expanding out the new $Q$ terms again using \eqref{eq:qexplicit}, we obtain the most explicit form of $G_N^{(m)}$:
\begin{equation} \label{eq:Gmnfinalcomp1}
    G^{(m)}_N(\textbf{g}) = \frac{(-1)^m}{m!} \sum_{\substack{A \in 2^\N, B \in 2^{\N \cup \{ 0 \}} \\ |A| = |B| = m}} \sum_{\alpha, \sigma, \tau \in S_m} (-1)^{(\sigma\inv \circ \tau)^A} E(w_k, z_k) \big|_{w_i^{-N - A_{\sigma_i} - B_{\alpha_i}} z_i^{N + A_{\tau_i} + B_{\alpha_i}}}.
\end{equation}
and if we wished to make both $A$ and $B$ vary over sets of positive integers only, we may re-express this as
\begin{equation} \label{eq:Gmnfinalcomp2}
    G^{(m)}_N(\textbf{g}) = \frac{(-1)^m}{m!} \sum_{\substack{A, B \in 2^\N \\ |A| = |B| = m}} \sum_{\alpha, \sigma, \tau \in S_m} (-1)^{\sigma} (-1)^{\tau} E(w_k, z_k) \big|_{w_i^{-N - A_{\sigma_i} - B_{\alpha_i} + 1} z_i^{N + A_{\tau_i} + B_{\alpha_i} - 1}},
\end{equation}
where $E(w_k,z_k)$ is given in (\ref{eq:Edef}).  

Equation \eqref{eq:Gmnfinalcomp2} is a significant simplification of our original expression, equation (\ref{eq:GmNfinal}). It provides an explicit way to identify $G^{(m)}_N(\textbf{g})$ as a sum over different power series coefficients of $E(w_k, z_k)$, and also cuts out many unnecessary terms that vanish due to the determinant structure of (\ref{eq:GmNfinal}).

\subsubsection{The order of $G_N^{(m)}(\mathbf g)$}
\label{sec:gglower}

With the original determinant expression, (\ref{eq:GmNfinal}), reexpressed as a sum over sets of distinct positive integers $A$ and $B$ in (\ref{eq:Gmnfinalcomp2}), we can now provide a lower bound on the order of $G_N^{(m)}(\textbf{g})$.  The general idea is that, for a given rank $N$ and number of giant gravitons, $m$, we are selecting terms corresponding to selected powers of $w_i$ and $z_i$ in the exponential factor $E(w_k,z_k)$ defined in (\ref{eq:Edef}).

As indicated at the beginning of this section, we assume the scaling $\gamma_k\sim q^{\alpha k}$.  When applied to the index, this corresponds to the single letter index $i(q)$ having order $\alpha$.  We now associate these powers of $q$ with $\alpha_k$ in (\ref{eq:Edef}), so that
\begin{equation}
    \log E(w_k,z_k)\sim \sum_{k=1}^\infty  \left( \sum_{i = 1}^m (q^{\alpha} z_i)^k - \sum_{i = 1}^m (q^{\alpha} w_i)^k \right) \left( \sum_{i = 1}^m z_i^{-k} - \sum_{i = 1}^m w_i^{-k} \right),
\end{equation}
where we are only concerned with the leading order terms in the series expansion in $q$, and where we have ignored constant factors.  In this way of writing the exponential factor, the order of $q$ is tied to positive powers of $z_i$.  We now recall from (\ref{eq:Gmnfinalcomp2}) that we are selecting positive powers of $z_i$ and negative powers of $w_i$.  Thus, if we wish to minimize the order of $q$, we should make this selecting as efficient as possible.  In particular, we should not pick any negative powers of $z_i$ nor any positive powers of $w_i$.  As a result, for power counting purposes, we can restrict to the expression
\begin{equation}
    \log E(w_k,z_k)\sim \sum_{k=1}^\infty  \left( \sum_{i = 1}^m (q^{\alpha} z_i)^k  \right) \left(\sum_{i = 1}^m w_i^{-k} \right).
\end{equation}
What this indicates is that we can count powers of $q$ in the same way we count powers of $z_i$.  (Note that all $z_i$'s scale identically with $q$.)

From (\ref{eq:Gmnfinalcomp2}), we need to consider each $z_i$ having power $N + A_{\tau_i} + B_{\alpha_i} - 1$.   Since we have multiple $z_i$'s, we must take the sum of these powers of the $z_i$ in order to get the total contribution to the order of $q$. Thus, for a given $A$ and $B$, the lower bound on the order is given by
\begin{equation} \label{eq:lowerboundAB1}
    \sum_{i = 1}^m (N + A_{\tau_i} + B_{\alpha_i} - 1)
    = mN + \sum_{i = 1}^m A_i + \sum_{i = 1}^m B_i - m.
\end{equation}
To get the true lower bound on the order of $G_N^{(m)}(\mathbf q)$, we note that the choice of $A$ and $B$ that minimizes the bound in \eqref{eq:lowerboundAB1} is $A = B = \{ 1,2, \dots, m \}$, and thus, the minimal order of the series is given by $G_N^{(m)}(\mathbf q)\sim q^{\alpha\mathbf{k}_{\mathrm{min}}}$ where

\begin{equation} \label{eq:Gmnlowerbound}
    \mathbf{k}_{\mathrm{min}} \geq m(N + m).
\end{equation}
A more rigorous proof of this lower bound is provided in Appendix~\ref{app:orderGmN}.

\section{Examples}
\label{sec:examples}

Our main result is an explicit formula, (\ref{eq:GmNfinal}), for the expansion of the matrix integral (\ref{eq:matint}).  This formula can also be rewritten in terms of a sum over partitions, (\ref{eq:Gmnfinalcomp2}), which is useful in determining the order of the terms in the expansion.  While this is a completely general result, the primary motivation is to apply this expansion to the $\mathcal N=4$ SYM index, (\ref{eq:ZNq}).  Along these lines, recall that the couplings $g_k$ in (\ref{eq:matint}) are to be identified with the single letter index, $g_k=i(q^k)$, where we consider a single fugacity $q$ for simplicity.

To compute the index, note that the effective couplings, $\gamma_k$, defined in (\ref{eq:gammak}), are given by $\gamma_k=\hat i(q)$ where
\begin{equation}
    \hat i(q)\equiv\fft{i(q)}{1-i(q)}.
\end{equation}  
The exponential factor in (\ref{eq:GmNfinal}) is essentially a plethystic exponential of $\gamma_k\alpha_k\beta_k$.  Following \cite{Murthy:2022ien}, we take the series expansion
\begin{equation}
    \hat i(q)=\sum_{n=1}^\infty\hat a_nq^n.
\end{equation}
We can then  perform the sum in the exponential to arrive at the giant graviton expression
\begin{align}
    G_N^{(m)}&=\left[
    \vphantom{\left(\fft{(1-\fft{z_i}{z_j}q^k)}{(1-\fft{z_i}{w_j}q^k)}\right)^{\hat a_k}}
    \fft{(-1)^m}{m!}\left(\prod_{i=1}^m\fft{w_i/z_i}{1-w_i/z_i}\right)\det\left(\fft1{1-w_j/z_i}\right)_{i,j=1}^m\right.\nn\\
    &\kern10em\times\left.\prod_{i,j=1}^m\prod_{k=1}^\infty\left(\fft{(1-\fft{z_i}{z_j}q^k)(1-\fft{w_i}{w_j}q^k)}{(1-\fft{z_i}{w_j}q^k)(1-\fft{w_i}{z_j}q^k)}\right)^{\hat a_k}\right]_{w_i^{-N}z_i^N}.
\label{eq:gginm}
\end{align}
This can be generalized to multiple fugacities if desired.  We now turn to some examples.

\subsection{The 1/2 BPS index}

For the 1/2 BPS $\mathcal N=4$ SYM index, the single letter index is $i(q)=q$, which gives
\begin{equation}
    \hat f(q)=\fft{q}{1-q}=\sum_{n=1}^\infty q^n.
\end{equation}
The index is easily evaluated, and gives $Z_N=\prod_{k=1}^N(1-q^k)^{-1}$, so that
\begin{equation}
    \fft{Z_N}{Z_\infty}=\prod_{k=1}^\infty(1-q^{N+k}).
\end{equation}
This indicates that the first finite $N$ correction to $Z_\infty$ occurs at $\mathcal O(q^{N+1})$, which matches the order of $G_N^{(1)}$.  Evaluation of the $m$ giant graviton contribution, $G_N^{(m)}$, is straightforward, and we give, as an example, the expansion of the index for $N=2$ in Table~\ref{tbl:1/2BPS}.  As expected, the contribution of $m$ giant gravitons starts at order $m(N+m)$.  However, it is curious to note that the $G_N^{(m)}$ expansion differs from the analytic result \cite{Gaiotto:2021xce,Murthy:2022ien}
\begin{equation}
    \fft{Z_N}{Z_\infty}=\sum_{m=0}^\infty\mathcal G_N^{(m)},\qquad\mbox{where}\qquad\mathcal G_N^{(m)}=(-1)^mq^{mN+m(m+1)/2}\fft1{\prod_{k=1}^m(1-q^k)}.
\label{eq:calG}
\end{equation}
For comparison, this expansion is also shown in Table~\ref{tbl:1/2BPS}.  Note that here the order of $\mathcal G_N^{(m)}$ is $mN+m(m+1)/2$, which differs from that of $G_N^{(m)}$.  This highlights the fact that the expansion of a series in terms of a set of ``giant graviton'' contributions is hardly unique, and that perhaps additional physics input may be required to unambiguously make a connection with giant gravitons in any supergravity dual.  This non-uniqueness also manifests itself in the 1/16 BPS index, to which we turn to next.

\begin{table}[t]
\begin{tabular}{l|rrrrrrrrrrrrrrrrr}
&$q^0$&$q^1$&$q^2$&$q^3$&$q^4$&$q^5$&$q^6$&$q^7$&$q^8$&$q^9$&$q^{10}$&$q^{11}$&$q^{12}$&$q^{13}$&$q^{14}$&$q^{15}$&$\cdots$\\
\hline$Z_2/Z_\infty$&$1$&&&$-1$&$-1$&$-1$&$-1$&&&$1$&$1$&$2$&$1$&$2$&$1$&$1$&$\cdots$\\
\hline
$G_2^{(1)}$&&&&$-1$&$-1$&$-1$&$-1$&&$-1$&&$-1$&&$-2$&&$-1$&$-1$&$\cdots$\\
$G_2^{(2)}$&&&&&&&&&$1$&$1$&$2$&$2$&$3$&$2$&$2$&$3$&$\cdots$\\
$G_2^{(3)}$&&&&&&&&&&&&&&&&$-1$&$\cdots$\\
\hline
$\mathcal G_2^{(1)}$&&&&$-1$&$-1$&$-1$&$-1$&$-1$&$-1$&$-1$&$-1$&$-1$&$-1$&$-1$&$-1$&$-1$&$\cdots$\\
$\mathcal G_2^{(2)}$&&&&&&&&$1$&$1$&$2$&$2$&$3$&$3$&$4$&$4$&$5$&$\cdots$\\
$\mathcal G_2^{(3)}$&&&&&&&&&&&&&$-1$&$-1$&$-2$&$-3$&$\cdots$
\end{tabular}
\caption{The 1/2 BPS index for $N=2$ and the giant graviton expansion with $G_N^{(m)}$ given in (\ref{eq:GmNfinal}).  Here we also present the alternate expansion in terms of $\mathcal G_N^{(m)}$ given in (\ref{eq:calG}).}
\label{tbl:1/2BPS}
\end{table}

\subsection{The 1/16 BPS index}

Much of the focus of the giant graviton expansion lies in the evaluation of the 1/16 BPS index.  Following  \cite{Kinney:2005ej,Romelsberger:2005eg}, and using the convention of \cite{Benini:2018ywd}, but with $Q_a^{\mathrm{here}}=\fft12R_a^{\mathrm{there}}$, the 1/16 BPS superconformal index is given by
\begin{equation}
\mathcal I(p,q;y_a)=\Tr\left[(-1)^Fe^{-\beta\{\mathcal Q,\mathcal Q^\dagger\}}p^{J_1}q^{J_2}y_1^{Q_1}y_2^{Q_2}y_3^{Q_3}\right],
\label{eq:WI}
\end{equation}
under the constraint $pq=y_1y_2y_3$.  With these conventions, the single letter index takes the form
\begin{equation}
    i(p,q;y_i)=1-\fft{(1-y_1)(1-y_2)(1-y_3)}{(1-p)(1-q)},
\end{equation}
so that
\begin{equation}
    \hat i(p,q;y_i)=\fft{(1-p)(1-q)}{(1-y_1)(1-y_2)(1-y_3)}-1.
\end{equation}
Note that the infinite $N$ index, corresponding to the multi-graviton contribution, takes the form
\begin{equation}
    Z_\infty=\prod_{k=1}^\infty\fft{(1-p^n)(1-q^n)}{(1-y_1^k)(1-y_2^k)(1-y_3^k)}.
\end{equation}
For simplicity, we take a single fugacity by setting
\begin{equation}
    p=q=\hat q^3,\qquad y_1=y_2=y_3=\hat q^2,
\label{eq:onefug}
\end{equation}
and then dropping the hat on $\hat q$ to obtain $i(q)=1-(1-q^2)^3/(1-q^3)^2$, as investigated in \cite{Imamura:2021ytr,Murthy:2022ien}.  Working out (\ref{eq:GmNfinal}) for $N=2$ then gives the expansion in terms of the giant graviton contributions $G_N^{(m)}$ shown in Table~\ref{tbl:N=2}.  Since in this case $i(q)$ starts at order $q^2$, the giant graviton expansion starts at order $2m(N+m)$ for $m$ giant gravitons.

\begin{table}[t]
\begin{tabular}{l|r|rrr|rrr}
&$Z_2/Z_\infty$&$G_2^{(1)}$&$G_2^{(2)}$&$G_2^{(3)}$&$\mathcal I_2^{(1)}/Z_\infty$&$\mathcal I_2^{(2)}/Z_\infty$&$\mathcal I_2^{(3)}/Z_\infty$\\
\hline
$q^0$&1&&&\\
$q^1$&&&&\\
$q^2$&&&&\\
$q^3$&&&&\\
$q^4$&&&&\\
$q^5$&&&&\\
$q^6$&$-10$&$-10$&&&$-10$\\
$q^7$&$12$&$12$&&&$12$\\
$q^8$&$-9$&$-9$&&&$-9$\\
$q^9$&&&&\\
$q^{10}$&$21$&$21$&&&$21$\\
$q^{11}$&$-54$&$-54$&&&$-54$\\
$q^{12}$&$83$&$83$&&&$83$\\
$q^{13}$&$-102$&$-102$&&&$-102$\\
$q^{14}$&$72$&$72$&&&$72$\\
$q^{15}$&$128$&$128$&&&$128$\\
$q^{16}$&$-459$&$-585$&$126$&&$-711$&$252$\\
$q^{17}$&$744$&$1122$&$-378$&&$1962$&$-1218$\\
$q^{18}$&$-697$&$-1513$&$816$&&$-4418$&$3721$\\
$q^{19}$&$12$&$1380$&$-1368$&&$9018$&$-9006$\\
$q^{20}$&$1440$&$138$&$1302$&&$-17145$&$18585$\\
$q^{21}$&$-3240$&$-3900$&$660$&&$30902$&$\textcolor{blue}{-34142}$\\
$q^{22}$&$4182$&$9996$&$-5814$&&$-53619$&$\textcolor{blue}{57801}$\\
$q^{23}$&$-2580$&$-17376$&$14796$&&$90090$&$\textcolor{blue}{-92670}$\\
$q^{24}$&$-2971$&$22568$&$-25539$&&$-147243$&$\textcolor{blue}{144272}$\\
$q^{25}$&$12132$&$-18114$&$30246$&&$235494$&$\textcolor{blue}{-223362}$\\
$q^{26}$&$-20220$&$-6030$&$-14190$&&$-369666$&$\textcolor{blue}{349446}$\\
$q^{27}$&$18118$&$58474$&$-40356$&&$569932$&$\textcolor{blue}{-551814}$\\
$q^{28}$&$2526$&$-142020$&$144546$&&$-864885$&$\textcolor{blue}{867411}$\\
$q^{29}$&$-41874$&$244116$&$-285990$&&$1295568$&$\textcolor{blue}{-1337442}$\\
$q^{30}$&$82815$&$-320713$&$405816$&$-2288$&$-1917827$&$\textcolor{blue}{2004503}$&$-3861$\\
$~\vdots$&$\vdots~~$&$\vdots~~$&$\vdots~~$&$\vdots~~$&$\vdots~~$&$\vdots~~$&$\vdots~~$
\end{tabular}
\caption{Comparison of the $1/16$ BPS index for $N=2$ with the giant graviton expansion, (\ref{eq:GmNfinal}), of \cite{Murthy:2022ien} for $G_N^{(m)}$ and the wrapped D3-brane expansion of \cite{Imamura:2021ytr} for $\mathcal I_N^{(m)}/Z_\infty$.  The single giant graviton expansions agree up to the point when two giant gravitons start to contribute.  (The numbers in blue are computed indirectly by subtracting $\mathcal I_2^{(1)}$ from $Z_2$, and hence do not provide an independent check of the expansion.)}
\label{tbl:N=2}
\end{table}

As indicated above for the 1/2 BPS case, the expansion of the index is not unique.  While the expression (\ref{eq:GmNfinal}), which provides an explicit formula for the expansion of \cite{Murthy:2022ien}, is certainly valid, it should be kept in mind that it was derived by direct manipulation of the matrix integral, (\ref{eq:ZNq}), corresponding to the field theory expression for the index.  Instead, if we took an AdS/CFT point of view, we could also approach the index from the supergravity side.  In this case, the giant gravitons have a direct interpretation as D3-branes wrapped on three-cycles inside $S^5$.  The $m$ giant graviton contribution can then be computed as the index of the worldvolume gauge theory of the wrapped D3-branes.  This was the approach taken in \cite{Arai:2019xmp,Imamura:2021ytr}, and the result yields a different expansion, also shown in Table~\ref{tbl:N=2} for comparison.

What we see is that the $G_N^{(m)}$ expansion of \cite{Murthy:2022ien} and the wrapped D3-brane expansion $\mathcal I_N^{(m)}$ of \cite{Imamura:2021ytr} are not the same.  In both cases, the $m$ giant graviton contribution starts at order $2m(N+m)$.  However, the expansion coefficients differ once the two-giant graviton contribution starts.  Given the structure of the expansion, there is no room for any difference when only the one-giant graviton contribution is present.  But once two or more giant gravitons show up we see the ambiguity in how to assign the expansion coefficients to the various giant graviton terms.

Although it is not necessarily surprising that the one-giant graviton contributions $G_2^{(1)}$ and $\mathcal I_2^{(1)}/Z_\infty$ start to differ once two giant gravitons show up, it is nevertheless curious that they actually agree for many terms (especially when $N$ is large) prior to deviating away from each other.  For a single giant graviton in the matrix model expansion, the result, (\ref{eq:gginm}), reduces to \cite{Murthy:2022ien}
\begin{equation}
    G_N^{(1)}=\left[-\fft\zeta{(1-\zeta)^2}\prod_{k=1}^\infty\left(\fft{(1-q^k)^2}{(1-\zeta q^k)(1-\zeta^{-1}q^k)}\right)^{\hat a_k}\right]_{\zeta^{-N}},
\label{eq:GN1}
\end{equation}
where $\zeta=w/z$, and $\hat a_k$ is the Taylor coefficient in the expansion of $\hat i=(1-q^3)^2/(1-q^2)^3-1$.  While we are unaware of a closed form expression for $G_N^{(1)}$, it can be evaluated up to reasonably high order on the computer.

On the other hand, using the index of the world volume theory on a single wrapped D3-brane, Refs.~\cite{Arai:2019xmp,Imamura:2021ytr} obtained the expression
\begin{equation}
    \fft{\mathcal I_N^{(1)}}{Z_\infty}=\mathcal I_{(1,0,0)}+\mathcal I_{(0,1,0)}+\mathcal I_{(0,0,1)},
\label{eq:D3index}
\end{equation}
where $\mathcal I_{(0,0,1)}=y_3^N\mbox{PE}(f_v^3)$, and the other contributions are obtained by cyclic rearrangement.  Here, the single letter index is
\begin{equation}
    f_v^3=1-\fft{(1-y_3^{-1})(1-p)(1-q)}{(1-y_1)(1-y_2)},
\end{equation}
and the regulated plethystic exponential takes the explicit form
\begin{equation}
    \mbox{PE}(f_v^3)=-\fft{y_3}{(1-y_1/y_3)(1-y_2/y_3)}\mbox{PE}\left(f_v^3-y_3^{-1}+y_3-y_1/y_3-y_2/y_3\right).
\label{eq:PEreg}
\end{equation}
Note that we have translated the notation of \cite{Arai:2019xmp,Imamura:2021ytr} to correspond to the fugacities given in (\ref{eq:WI}).  In order to reduce this to a single fugacity, we cannot directly substitute (\ref{eq:onefug}) into the expression for $\mathcal I_{(0,0,1)}$ because the denominators in (\ref{eq:PEreg}) would then vanish.  However, these denominators cancel in the combination (\ref{eq:D3index}), and the resulting expressions can be evaluated on the computer.  The entries for $\mathcal I_2^{(1)}$ in Table~\ref{tbl:N=2} were obtained from (\ref{eq:D3index}), while the entries for $\mathcal I_2^{(2)}$ and $\mathcal I_2^{(3)}$ were obtained from \cite{Imamura:2021ytr}, with the exception of the terms in blue, which were obtained implicitly by subtracting $\mathcal I_2^{(1)}$ from $Z_2$.

The two expressions, (\ref{eq:GN1}) for $G_N^{(1)}$ and (\ref{eq:D3index}) for $\mathcal I_N^{(1)}/Z_\infty$, were derived with entirely separate methods and in fact look rather distinct from each other.  Thus it is somewhat surprising that they agree in the series expansion starting at $q^{2(N+1)}$ up to $q^{4(N+2)-1}$, and then differ starting at $q^{4(N+2)}$ where the two-giant graviton contribution first shows up.  It would be curious to see if we could make a more direct connection between these two expansions.  To do so, it is likely we would have to better understand the contribution of two giant gravitons, which on the supergravity side partially involves the worldvolume theory of two coincident D3-branes.  The index of the worldvolume theory of coincident branes has been investigated in \cite{Imamura:2022aua}.

\section{Discussion}
\label{sec:discussion}

As we have seen, the $\mathcal N=4$ SYM index can be expanded in terms of a `giant graviton expansion' where the contribution of $m$ giant gravitons starts at $\mathcal O(q^{\alpha m(N+m)})$ for a fixed integer $\alpha$ corresponding to the order of the single letter index $i(q)$.  However, the expansion is not unique once two or more giant gravitons can contribute.  From a holographic point of view, the natural expansion would be that of wrapped D3-branes \cite{Arai:2019xmp,Imamura:2021ytr}.  The expansion of the giant graviton index then involves a contribution of indices from the worldvolume theories of wrapped D3-branes
\begin{equation}
    \fft{Z_N}{Z_\infty}=\sum_{n_1,n_2,n_3}\mathcal I_{(n_1,n_2,n_3)},
\end{equation}
where the $n_i$ are wrapping numbers on the different three-cycles inside $S^5$.  From this point of view, the $m$ giant graviton contribution $\mathcal I_N^{(m)}/Z_\infty$ is a sum of indices of D3-branes with total wrapping number $m=n_1+n_2+n_3$.  However, this sum over wrapping numbers on the three distinct cycles is absent in the $G_N^{(m)}$ expansion of the matrix integral in the approach of \cite{Murthy:2022ien}.  From this point of view, it is perhaps not surprising that the $G_N^{(m)}$ and $\mathcal I_N^{(m)}$ expansions differ.  At the same time, this makes us wonder whether it would be possible to modify the expansion of \cite{Murthy:2022ien} to make it more closely resemble the wrapped D3-brane expansion and hence to find a direct match between the matrix integral and supergravity expansions of the index.  In order to do so, one would probably have to specialize away from the generic matrix integral, (\ref{eq:matint}), and to focus directly on $\mathcal N=4$ SYM, where the D3-brane wrapping cycles are identified with the maximal torus of the R-symmetry group SU(4)$_R$.

Finally, as an expansion in powers of $q$, at $\mathcal O(1)$ we only see the multigraviton contribution, while at $\mathcal O(N)$ the giant gravitons start to contribute.  Pushing to higher orders, we can wonder if anything qualitatively changes at $\mathcal O(N^2)$, which is the black hole regime in the supergravity dual \cite{Cabo-Bizet:2018ehj,Choi:2018hmj,Benini:2018ywd}.  Of course, the expansion (\ref{eq:gge}) is valid to arbitrarily high powers of $q$.  But in the large-$N$ limit one would expect on the order of $N$ wrapped D3-branes to strongly backreact on the geometry, leading to a more natural black hole description of the index, at least from the supergravity point of view.  Such complications also arise in the Cardy-like limit when $q\to1$ since this approaches the radius of convergence of the various series expansions in $q$ used throughout the manipulations of the matrix integral.

It is thus clear that much remains to be understood in connecting the various giant graviton expansions with each other and with the supergravity dual.  Nevertheless, we already see hints of intricate structures in the finite-$N$ index.  Ultimately, by further study of the index, we may hope to develop a closer connection between superconformal gauge theories and their gravity duals, especially at finite $N$ where quantum gravity effects can no longer be ignored.

\section*{Acknowledgements}

JTL wishes to thank S.\ Murthy for enlightening discussions.  NJR was supported through a University of Michigan Physics Summer Research Opportunity award.  This work was supported in part by the U.S. Department of Energy under grant DE-SC0007859.

\appendix

\section{Rewriting the giant graviton expression}
\label{app:gmnrew}

In this appendix, we elaborate on the computation in Section~\ref{sec:ggeorder} where we rewrite the expression for the contribution of $m$ giant gravitons, \eqref{eq:GmNfinal}, in terms of the expanded $P(k_1,\ldots,k_m)$ terms of \eqref{eq:detterms}.  For ease of notation, we denote the exponential term in (\ref{eq:GmNfinal}) by $E(w_k, z_k)$, as given in (\ref{eq:Edef}). Expanding the $(w_i/z_i)/(1 - w_i/z_i)$ prefactor in (\ref{eq:GmNfinal}) as a power series in $w_i/z_i$ as well as the determinant according to \eqref{eq:detexpr}, we obtain
\begin{equation} \label{eq:simpstep1}
    \hat K_m (\textbf{g}) = \frac{(-1)^m}{m!} \cdot \prod_{j = 1}^m \left( \sum_{\ell_j = 0}^\infty \left( \frac{w_j}{z_j} \right)^{\ell_j} \right) \cdot \left( \prod_{i = 1}^m \frac{w_i}{z_i} \sum_{\substack{A \in 2^{\N \cup \{0\}} \\ |A| = m}} \sum_{\tau \in S_m} P(A, \tau) \right) \cdot E(w_k, z_k).
\end{equation}
Going back to the definition of $P(A, \tau)$ as in \eqref{eq:detterms} and using the multilinearity of the determinant, we see that
\begin{align}
    \prod_{i = 1}^m \frac{w_i}{z_i} \sum_{\substack{A \in 2^{\N \cup \{0\}} \\ |A| = m}} \sum_{\tau \in S_m} P(A, \tau) 
    &= \sum_{\substack{A \in 2^{\N \cup \{0\}} \\ |A| = m}} \sum_{\tau \in S_m} \det \left( \left( \frac{w_j}{z_i} \right)^{A_{\tau_i} + 1} \right)_{i, j = 1}^m.
\end{align}
But here, note that $A_{\tau_i} + 1 = (A + 1)_{\tau_i}$. This is because we add 1 to every term in $A$, and this has no effect on permutating the actual elements of $A$.  We thus arrive at
\begin{align}
    \prod_{i = 1}^m \frac{w_i}{z_i} \sum_{\substack{A \in 2^{\N \cup \{0\}} \\ |A| = m}} \sum_{\tau \in S_m} P(A, \tau)
    &= \sum_{\substack{A \in 2^{\N \cup \{0\}} \\ |A| = m}} \sum_{\tau \in S_m} \det \left( \left( \frac{w_j}{z_i} \right)^{(A + 1)_{\tau_i}} \right)_{i, j = 1}^m \nn\\
    &= \sum_{\substack{A \in 2^{\N \cup \{0\}} \\ |A| = m}} \sum_{\tau \in S_m} P(A + 1, \tau) \nn\\
    &= \sum_{\substack{A \in 2^\N \\ |A| = m}} \sum_{\tau \in S_m} P(A, \tau).
\end{align}
Plugging our result back into \eqref{eq:simpstep1} and rearranging terms gives us \eqref{eq:Kmint1}.

\section{The lower bound on the order of $G_N^{(m)}(\textbf{g})$}
\label{app:orderGmN}

This appendix furnishes a more rigorous proof of the order of the contributions of $m$ giant gravitons, (\ref{eq:Gmnlowerbound}), given in Section~\ref{sec:gglower}.  As in Section \ref{sec:ggeorder}, we assume $\gamma_k \sim q^{\alpha k}$. Also, note that since we are simply looking for the order, we may freely ignore numerical coefficients for convenience. We note that the individual terms in \eqref{eq:Gmnfinalcomp2} are given by
\begin{align} \label{eq:Gmnfinalterms}
    & \left[ \exp \left( - \sum_{k = 1}^\infty \frac{\gamma_k}{k} \left( \sum_{i = 1}^m z_i^k - \sum_{i = 1}^m w_i^k \right) \left( \sum_{i = 1}^m z_i^{-k} - \sum_{i = 1}^m w_i^{-k} \right) \right) \right]_{w_i^{-N - A_{\sigma_i} - B_{\alpha_i} + 1} z_i^{N + A_{\tau_i} + B_{\alpha_i} - 1}} \\
    &\sim \left[ \sum_{n = 0}^\infty \frac{1}{n!} \left( - \sum_{k = 1}^\infty \frac{q^{\alpha k}}{k} \left( \sum_{i = 1}^m z_i^k - \sum_{i = 1}^m w_i^k \right) \left( \sum_{i = 1}^m z_i^{-k} - \sum_{i = 1}^m w_i^{-k} \right) \right)^n \right]_{w_i^{-N - A_{\sigma_i} - B_{\alpha_i} + 1} z_i^{N + A_{\tau_i} + B_{\alpha_i} - 1}}.
\end{align}
Now, note that every term that results from the product
\begin{equation}
    q^{\alpha k} \left( \sum_{i = 1}^m z_i^k - \sum_{i = 1}^m w_i^k \right) \left( \sum_{i = 1}^m z_i^{-k} - \sum_{i = 1}^m w_i^{-k} \right),
\end{equation}
(modulo numerical coefficients) has the form
\begin{equation}
    q^{\alpha k} z_{p_1}^{k a_1} z_{p_2}^{-k a_2} w_{p_3}^{k a_3} w_{p_4}^{-k a_4},
\end{equation}    
where $a_o \in \{ 0, 1 \}$ for $o = 1, \dots, 4$ and exactly two of them are nonzero. This means that the terms that comprise the expression
\begin{equation}
    \sum_{n = 0}^\infty \frac{1}{n!} \left( - \sum_{k = 1}^\infty \frac{q^{\alpha k}}{k} \left( \sum_{i = 1}^m z_i^k - \sum_{i = 1}^m w_i^k \right) \left( \sum_{i = 1}^m z_i^{-k} - \sum_{i = 1}^m w_i^{-k} \right) \right)^n,
\end{equation}
must have the form (modulo numerical coefficients)
\begin{equation} \label{eq:genterm}
    q^{\alpha \textbf{k}} \prod_{\ell=1}^n z_{p_1, \ell}^{k_\ell a_{1,\ell}} \prod_{\ell=1}^n z_{p_2, \ell}^{-k_\ell a_{2,\ell}} \prod_{\ell=1}^n w_{p_3, \ell}^{k_\ell a_{3,\ell}} \prod_{\ell=1}^n w_{p_4, \ell}^{-k_\ell a_{4,\ell}},
\end{equation}
where
\begin{equation}
    \textbf{k} = \sum_{\ell = 1}^n k_\ell.
\end{equation}
As before, $a_{o,\ell} \in \{ 0, 1 \}$ for $1 \leq j \leq 4$ and $1 \leq \ell \leq n$, and for a given $\ell$, exactly two of the $a_{o, \ell}$ are nonzero.

Now, note that for all $1 \leq o \leq 4$ and $1 \leq \ell \leq n$, the $z_{p_o, \ell}$ are just equal to some $z_i$ for $1 \leq i \leq m$. In view of this, we would like to reorganize \eqref{eq:genterm} in terms of the $z_i$'s. To this end, for $1 \leq i \leq m$, let us define $\ell_{o, i,1}, \dots, \ell_{o, i, n_{o,i}} \in \N$ such that $p_{o, \ell_{o, i, j}} = i$ for all $1 \leq o \leq 4$ and $1 \leq j \leq n_{o,i}$. Through this reorganization, we may rewrite
\begin{equation}
    \prod_{\ell=1}^n z_{p_1, \ell}^{k_\ell a_{1, \ell}} = \prod_{i=1}^n z_i^{\sum_{j = 1}^{n_{1,i}} k_{\ell_{1, i,j}} a_{1, \ell_{1,i,j}}},
\end{equation}
and obtain similar expressions for the other terms. Combining these results together, \eqref{eq:genterm} may be re-expressed as
\begin{equation}
    q^{\alpha \textbf{k}} \prod_{i=1}^n z_i^{\bm \sigma_i} w_i^{\bm \rho_i},
\end{equation}
where
\begin{equation} \label{eq:sigmaorder}
    \bm \sigma_i = \sum_{j = 1}^{n_{1,i}} k_{\ell_{1, i,j}} a_{1, \ell_{1,i,j}} - \sum_{j = 1}^{n_{2,i}} k_{\ell_{2, i,j}} a_{2, \ell_{2,i,j}},\qquad
    \bm \rho_i = \sum_{j = 1}^{n_{3,i}} k_{\ell_{3, i,j}} a_{3, \ell_{3,i,j}} - \sum_{j = 1}^{n_{4,i}} k_{\ell_{4, i,j}} a_{4, \ell_{4,i,j}}.
\end{equation}
Going back to \eqref{eq:Gmnfinalterms}, we recall that we wish to select the term $$\prod_{i = 1}^m w_i^{-N - A_{\sigma_i} - B_{\alpha_i} + 1} z_i^{N + A_{\tau_i} + B_{\alpha_i} - 1}$$ This corresponds to setting
\begin{equation}
    \bm \sigma_i = -\bm \rho_i = N + A_{\tau_i} + B_{\alpha_i} - 1.
\end{equation}
Searching for a lower bound on the order of $q$ now translates to finding a lower bound for $\textbf{k}$, which we can now rewrite as
\begin{equation}
    \textbf{k} = \sum_{i = 1}^m \sum_{j = 1}^{n_{1,i}} k_{\ell_{1,i,j}}.
\end{equation}
We may replace the 1 in the indices with 2, 3, or 4 and this would still hold, since they are all just different partitions of the same indices. For our purposes though, the above is sufficient.

Now, noting that $a_\ell \in \{ 0, 1 \}$, $k_\ell \in \N$, and the constraints from \eqref{eq:sigmaorder}, we see that
\begin{equation}
    \sum_{j = 1}^{n_{1,i}} k_{\ell_{1,i,j}} \geq \sum_{j = 1}^{n_{1,i}} k_{\ell_{1, i,j}} a_{\ell_{1,i,j}} 
    = \bm \sigma_i + \sum_{j = 1}^{n_{2,i}} k_{\ell_{2, i,j}} a_{\ell_{2,i,j}}
    \geq \bm \sigma_i,
\end{equation}
which means that
\begin{equation} \label{eq:lowerboundAB}
    \textbf{k} = \sum_{i = 1}^m \sum_{j = 1}^{n_{1,i}} k_{\ell_{1,i,j}} \geq \sum_{i = 1}^m \bm \sigma_i
    = \sum_{i = 1}^m (N + A_{\tau_i} + B_{\alpha_i} - 1)
    = mN + \sum_{i = 1}^m A_i + \sum_{i = 1}^m B_i - m.
\end{equation}
In fact, this lower bound is saturated: setting $a_{1, \ell_{1,i,j}} = a_{3, \ell_{3,i,j}} = 1$ and $a_{2,\ell_{2,i,j}} = a_{4, \ell_{4,i,j}} = 0$ does the trick.

To get the true lower bound, we note that the choice of $A$ and $B$ that minimizes the bound in \eqref{eq:lowerboundAB} is $A = B = \{ 1,2, \dots, m \}$, and thus, we have that the minimal order of the series is given by
\begin{equation}
    \alpha \textbf{k}_{min} \geq \alpha \left( mN + 2 \cdot \frac{m(m + 1)}{2} - m \right)
    = \alpha m(N + m).
\end{equation}

\bibliographystyle{JHEP}
\bibliography{references}

\end{document}